\documentclass[aps,showpacs,pra,10pt,superscriptaddress,twocolumn]{revtex4}
\usepackage{mathrsfs}

\usepackage{amsmath,amsfonts,amssymb,graphics,graphicx,epsfig,color,times,indentfirst,layout}

\begin{document}

\title{Simulation of the spin-boson model with superconducting phase qubit coupled to a transmission line}

\author{Long-Bao Yu}

\affiliation{Laboratory of Quantum Information Technology, ICMP
and SPTE, South China Normal University, Guangzhou, China}
\affiliation{Department of Physics and Electronic Engineering,
Hefei Normal University, Hefei, China}

\author{Ning-Hua Tong}

\affiliation{Department of Physics, Renmin University of China,
Beijing, China}

\author{Zheng-Yuan Xue}

\affiliation{Laboratory of Quantum Information Technology, ICMP
and SPTE, South China Normal University, Guangzhou, China}

\author{Z. D. Wang}

\affiliation{Department of Physics and Center of Theoretical and
Computational Physics, The University of Hong Kong, Pokfulam Road,
Hong Kong, China}

\author{Shi-Liang Zhu}
\email{slzhu@scnu.edu.cn} \affiliation{Laboratory of Quantum
Information Technology, ICMP and SPTE, South China Normal
University, Guangzhou,  China}

\begin{abstract}
Based on the rapid  experimental developments  of circuit QED, we
propose a feasible scheme to simulate a spin-boson model with the
superconducting circuits, which can be used to detect quantum
Kosterlitz-Thouless (KT) phase transition. We design the
spin-boson model by using a superconducting phase qubit coupled
with a semi-infinite transmission line, which is regarded as
bosonic reservoir with a continuum spectrum. By tuning the bias
current or the coupling capacitance, the quantum KT transition can
be directly detected through tomography measurement on the states
of the phase qubit. We also estimate the experimental parameters
using numerical renormalization group method.
\end{abstract}

\pacs{85.25.Cp, 74.40.Kb, 03.67.Ac, 05.10.Cc}

\maketitle %
%\section{introduction}
Quantum simulation is one of the original inspirations for quantum
computing, proposed by Feynman \cite{Feynman} to solve the
difficulties of simulating quantum systems on a classical
computer. Simulating an arbitrary quantum system by the most
powerful classical computer is very hard for large scale quantum
systems, because of the exponential scaling of the Hilbert space
with the size of the quantum system. Quantum phase transitions
(QPT) \cite{Sachdev} at zero temperature play a key role in the
occurrence of important collective phenomena in quantum many-body
systems, which occur as a result of competing ground state phases.
Similarly, simulation of QPT with classical computer is also
difficulty since it is usually relevant to a quantum many-body
system.

On the other hand, the spin-boson model \cite{Leggett}, a
two-level system linearly coupled to a collective of harmonic
oscillators, is a typical model to study decoherence effects and
QPT. Those dissipative spin systems \cite{Weiss} are very
interesting because they display both a localized (classical) and
delocalized (quantum) phase for the spin. Spin-boson model has
been  primarily investigated by numerical renormalization group
(NRG) \cite{NRG, Bulla} and find that different spectral functions
of the bath may induce various kind of QPT. Many efforts have been
made to observe such environment-induced QPT in various systems,
such as mesoscopic metal ring \cite{Tong}, single-election
transistors \cite{Hur,Furusaki} and cold atoms \cite{Orth}, etc..

In this paper, we propose a feasible scheme to simulate the
spin-boson model with superconducting circuits, which can be used
to observe the notable quantum KT phase transition. Here we focus
on the ohmic case ($s =1$), which can be mapped on the anisotropic
Kondo model. The model shows a Kosterlitz-Thouless (KT) quantum
phase transition \cite{Leggett}, separating the localized phase at
$\alpha\geq\alpha_{c}$ from the delocalized phase at $\alpha
<\alpha_{c}$ \cite{Hur2}, where $\alpha$ represents the strength
of the dissipation and $\alpha_c$ is the critical value.
 Our idea is inspired by previous efforts to study the superconducting circuits as an artificial atom
 \cite{Makhlin,You,Yu,cQED,cQED2,Huang,Zhou}. In the paper we
design the spin-boson model using a superconducting phase qubit
coupled with a semi-infinite transmission line, which is regarded
as the qubit's environment with gapless spectra. In this setup,
the spectral function of the bosonic bath is treated as Ohmic. By
tuning the bias current or the coupling capacitance, the coupling
between the spin and the environment can be controlled. So the
states of the qubit may transit from delocalized phase to
localized state, which can be directly observed by measuring the
phase qubit. We also estimate the experimental parameters of this
transition with NRG method. Comparing with other candidates of
spin-boson model\cite{Tong,Hur,Furusaki,Orth}, the proposed
experimental setup based on superconducting system
\cite{Makhlin,You,Yu} may have some distinct advantages, such as
the parameters in the model are tunable through experimentally
controllable bias currents or driving microwaves, the bosonic
reservoir with a continuum spectrum can be simulated easily, and
the measurements are of high-fidelity, etc..

%\section{The experiment setup of spin-boson model}

\begin{figure}
\vspace{0.8cm}
\includegraphics[height=3.5cm,width=7.5cm,angle=0]{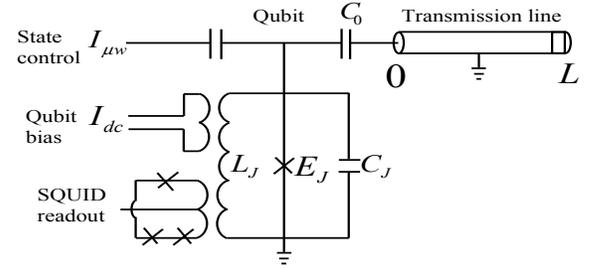}
\caption{A phase qubit capacitively coupled to a semi-infinite
transmission line simulates the spin-boson model.} \label{FIG.1}
\end{figure}

Our designed experimental setup is illustrated in Fig.\ref{FIG.1},
an approximately 1D transmission line (with the length as
$L\rightarrow\infty$) is capacitively coupled to  a current-biased
phase qubit. The phase qubit is an artificial spin, while  for
relatively low frequencies the transmission line is well described
by an infinite series of inductors with each node capacitively
connected to ground and then  can be considered as a boson
bath\cite{Huang}. The Hamiltonian of the whole system can be
written as
\begin{equation}
\label{H}
 H=H_{q}+H_{T}+H_{int},
\end{equation}
where
\begin{equation}
\label{Hq}
H_q=\frac{C_{J}}{2}(\frac{\Phi_{0}}{2\pi}\dot{\phi}_{J})^{2}-\frac{\Phi_{0}I_{b}}{2\pi}\phi_{J}-E_{J}\cos\phi_{J}
\end{equation}
is the Hamiltonian of the current-biased phase qubit,
\begin{equation}
\label{H_T}
H_T=\int^{L}_{0}\left\{\frac{1}{2c}\left[\frac{\partial\vartheta(x,t)}{\partial
x}\right]^{2}+\frac{l}{2}\dot{\vartheta}^{2}(x,t)\right\}dx
\end{equation}
is the Hamiltonian of the transmission line, and
\begin{equation}
\label{H_int} H_{int}=\int_0^L
\frac{C_{0}}{2}\left(\frac{1}{c}\frac{\partial\vartheta(0,t)}{\partial
 x}-\frac{\Phi_{0}}{2\pi}\dot{\phi}_{J}\right)^{2} dx
\end{equation}
is the interaction between the phase qubit and the transmission
line. In these equations $E_{J}=\frac{\Phi_{0}I_{0}}{2\pi}$ is the
magnitude of maximum Josephson coupling energy, $I_{0}$ is the
critical current of the junction, $\Phi_{0}=\frac{h}{2e}$ is the
superconducting flux quantum, $I_{b}$ is the bias current,
$\phi_{J}$ is the phase difference of the junction, $C_{J}$ and
$C_{0}$ are the junction and coupling capacitance, $c$ and $l$ are
the capacitance and inductance per unit length, respectively;
$\vartheta(x,t)=\int_{0}^{x}q(x',t)dx'$ is the collective charge
variable on the transmission line.

For the current-biased Josephson junction \cite{Martinis}, the
charge operator $\hat{Q}=C\frac{\Phi_{0}}{2\pi}\dot{\phi}_{J}$ and
phase difference operator $\hat{\phi}_{J}$ have the commutation
relationship $[\hat{\phi}_{J},\hat{Q}]=2ei$. Quantum mechanical
behavior can be observed for large area junctions in which
$E_{J}\gg E_{C}=e^{2}/2C$ when the bias current is slightly
smaller than the critical current $I_{b}\lesssim I_{0}$. In this
regime, the last two terms of the Hamiltonian of the phase qubit
(as Eq.(\ref{Hq})) can be accurately approximated by a cubic
potential $U(\phi_{J})$ parameterized by a barrier height $\Delta
U(I_{b})=(2\sqrt{2}I_{0}\Phi_{0}/3\pi)[1-(I_{b}/I_{0})]^{3/2}$ and
a quadratic curvature at the bottom of the well that gives a
classical oscillation frequency $\omega_{p}(I_{b})=2^{1/4}(2\pi
I_{0}/\Phi_{0}C)^{1/2}[1-(I_{b}/I_{0})]^{1/4}$  with the
capacitance $C=C_{J}+C_{0}$. The lowest two of the quantized
energy levels in the cubic potential are considered as the qubit
states, $\{|0\rangle,|1\rangle\}$, where
$\omega_{10}\simeq0.95\omega_{p}$ is the energy difference between
the ground state and the first excited state of the phase qubit.
The states can be fully manipulated with low- and microwave
frequency control currents, which can be chosen as
$I_{b}=I_{dc}+I_{\mu w}(t)$ \cite{Martinis}, so the phase qubit
can be expressed as
\begin{equation}\label{}
H_{q}=\frac{\epsilon}{2}\sigma_{x}-\frac{\Delta}{2}\sigma_{z},
\end{equation}
where $\sigma_{x,z}$ are Pauli matrices,
$\epsilon=\sqrt{\frac{\hbar}{2\omega_{10}C}}I_{\mu w}(t)$ and
$\Delta=\hbar\omega_{10}$. In the system, the charge operator can
be described as
$\hat{Q}=\sqrt{\frac{C\Delta}{2}}\hat{\sigma}_{y}$.

The corresponding Euler-Lagrange equation for the transmission line
is a wave equation
$$\frac{1}{c}\frac{\partial^{2}\vartheta(x,t)}{\partial
 x^{2}}-\frac{\vartheta(x,t)\delta(x)}{C_{0}}-\frac{l}{2}\frac{\partial^{2}\vartheta(x,t)}{\partial
 t^{2}}=0$$ with the mode speed $v=1/\sqrt{lc}$.
This is analogous to the problem of the Schr\"{o}dinger equation
with a delta function potential. By separation of variables, the
spatial part of the solution of the modes is of the form
\begin{equation}
\vartheta_{S}(x)=
\begin{cases}
A\cos(kx)& \text{symmetric modes},\\
A\sin(kx)& \text{antisymmetric modes},
\end{cases}
\end{equation}
where $0<x<L$. According to the boundary condition
$$\frac{1}{c}\frac{\partial\vartheta(0)}{\partial
 x}=\frac{\vartheta(L)}{C_{0}}$$
in the limit $C_{0}\ll c$, the spatial mode can be expressed as
%\begin{equation}
$k\approx \frac{m\pi}{2L},$
%\end{equation}
where $m$ is odd for symmetric modes and is even for antisymmetric
modes. The eigenfrequencies of the modes are $\omega=kv$. The time
dependent part for transmission line still has the form of
Euler-Lagrange equation
$$\ddot{\phi}_{m}(t)+\omega_{m}\phi_{m}(t)=0.$$ From the above
equation we can obtain the time dependent Hamiltonian for multi-mode transmission line
\begin{equation}\label{H(t)}
H_{T}(t)=\sum_{m}\frac{l}{2}\dot{\phi}^{2}_{m}(t)+\frac{1}{2c}(\frac{m\pi}{2L})^{2}\phi_{m}^{2}
\end{equation}
as a function of $\phi_{m}(t)$ and its canonically conjugate
momentum $p_{m}=l\dot{\phi}_{m}(t)$ with
$[\phi_{m},p_{m'}]=i\hbar\delta_{mm'}$.

To diagonalize the Hamilitonian $H_{T}(t)$,  we use the usual relations
\begin{subequations}
\begin{equation}
\hat{\phi}_{m}(t)=\sqrt{\frac{\hbar\omega_{m}c}{2}}\frac{2L}{m\pi}[a_{m}(t)+a^{\dagger}_{m}(t)],
\end{equation}
\begin{equation}
\hat{p}_{m}(t)=-i\sqrt{\frac{\hbar\omega_{m}l}{2}}[a_{m}(t)-a^{\dagger}_{m}(t)]
\end{equation}
\end{subequations}
by introducing the bosonic creation $a^{\dagger}$ and annihilation
operators $a$: $[a_{m}, a^{\dagger}_{m'}]=\delta_{mm'}$. Then the
collective charge variable in the semi-infinite transmission line is
%\begin{widetext}
%\begin{equation}
$\hat{\vartheta}(x,t)=\sum_{m=1}^{
m_{c}}\sqrt{\frac{\hbar\omega_{m}c}{L}}\frac{2L}{m\pi}[\hat{a}_{m}(t)+\hat{a}^{\dag}_{m}(t)]
\begin{cases}
\cos\frac{m\pi}{2L}x & \text{(m odd)},\\
\sin\frac{m\pi}{2L}x & \text{(m even)}.
\end{cases}$
%\end{equation}
%\end{widetext}
The voltage at the first end of the transmission line
($x=0$) is
\begin{equation}\label{H_int1}
\hat{V}(0,t)=\frac{1}{c}\frac{\partial\hat{\vartheta}(0,t)}{\partial
x}=\sum_{n=1}^{
n_{c}}\sqrt{\frac{\hbar\omega_{n}}{Lc}}[\hat{a}_{n}(t)+\hat{a}^{\dag}_{n}(t)],
\end{equation}
where $n=m/2=1,2,...n_{c}$. Here $m$ is even for the maximal
voltage. The voltage is zero for odd $m$ and  thus this part can
be neglected. Substituting Eq.(\ref{H_int1}) into $H_{int}$ in
Eq.(\ref{H_int}), we obtain the interaction Hamiltonian between
the phase qubit and the transmission line given by
\begin{equation}\label{}
H_{int}=C_{0}\sqrt{\frac{\Delta}{2C}}\sigma_{y}\sum_{n=1}^{
n_{c}}\sqrt{\frac{\hbar\omega_{n}}{Lc}}(a_{n}+a^{\dag}_{n}).
\end{equation}

After rotating the frame of the phase qubit ($\sigma_{z}\rightarrow\sigma_{x},\sigma_{y}\rightarrow\sigma_{z}$),
the Hamiltonian of the circuit without microwave frequency control current
($\epsilon=0$) can be written as the standard spin-boson model,
\begin{equation}\label{}
H=-\frac{\Delta}{2}\sigma_{x}+\hbar\omega_{n}a^{\dag}_{n}a_{n}+\frac{\sigma_{z}}{2}\sum_{n}\lambda_{n}(a^{\dag}_{n}+a_{n}),
\end{equation}
where $\lambda_{n}=C_{0}\sqrt{\frac{2\Delta\hbar\omega_{n}}{CLc}}$
are the coupling strengths between the spin and the modes of
transmission line, and the frequencies of modes
$\omega_{n}=\frac{n\pi}{L}\frac{1}{\sqrt{lc}}$. It is notable that
the coupling strengths can be controlled by the bias current $I_b$
(or the coupling capacitance $C_0$ if it can be modified in the
experiments), so the parameters in the spin-boson model realized
in this superconducting circuits are tunable and thus the system
is suitable to be used to observe rich phenomena in the spin-boson
model.

 The spectra is gapless when the length of the transmission
line is sufficient large. In this case, the bosonic bath
 can be characterized by its spectral function \cite{Leggett}
\begin{equation}\label{}
J(\omega)=\pi\sum_{n}\lambda_{n}^{2}\delta(\omega_{n}-\omega)=2\pi\alpha\omega^{s}\omega_{c}^{1-s}.
\end{equation}
In the proposed experimental setup, the obtained spectral function
is actually Ohmic case $s=1$. In additional, the dimensionless
parameter $\alpha=\frac{\Delta}{\pi}\frac{
C_{0}^{2}}{C}\sqrt{l/c}\simeq\frac{ C_{0}^{2}}{\pi
C}\hbar\omega_{p}\sqrt{l/c}$ measures the strength of the
dissipation, which is determined by the coupling strengths
$\lambda_n$ and may be modified with the bias current $I_{b}$ and
the coupling capacitance $C_{0}$.

%\section{numerical computation by NRG and discussion}

We now turn to address a method to observe one of the most
interesting phenomena in the spin-boson model: the KT phase
transition. In the dissipation case, the spin-boson model has a
delocalized and a localized zero temperature phases separated by
the KT transition \cite{Leggett}  at the critical value
$\alpha_{c}\simeq 1$ (for the unbiased case of $\epsilon=0$). In
the delocalized phase at small dissipation strength $\alpha$, the
ground state is nondegenerate and represents a damped tunneling
particle. For large $\alpha$, the dissipation leads to a
localization of the particle in one of the two $\sigma_{z}$
eigenstates, thus the ground state is doubly degenerate.

In the proposed experimental setup, the strength of the
dissipation is proportional to the controllable parameter as
$\alpha\propto\Delta/2\pi$, which can be changed by tuning the
bias current $I_{b}$.  The parameter $\alpha$ can be
experimentally modified in the regime $0.2\sim 3.0$ if one takes
the following typical data: the junction capacitance $C_{J}=0.85$
pF, $C_{0}=5C_{J}$, the impendence of transmission line
$z=\sqrt{l/c}=50$ $\Omega$, and $\Delta\simeq\hbar\omega_{p}/2$.
Therefore, the localized and delocalized states as well as the KT
phase transition may be observed by tuning the parameter
$\Delta/2$ through the biased current $I_b$.

\begin{figure}
\includegraphics[width=5.5cm, angle=90]{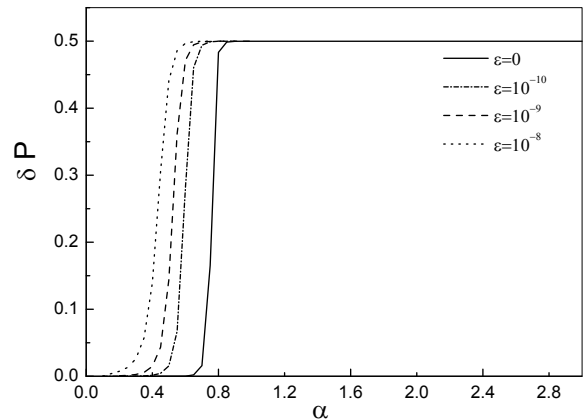}
\caption{The differences of the occupation probability $\delta
P=|P_{e}-P_{g}|/2$  for several values of $\epsilon$. The sudden
change of the populations characterizes the KT type quantum phase
transition in the spin-boson model, which occurs from the
delocalized phase at $\delta P=0$ to the localized phase at
$\delta P=\frac{1}{2}$.} \label{Fig2}
\end{figure}

The different phases and the KT phase transition can be
demonstrated through measuring the populations of the qubit states
$\langle\sigma_{z}\rangle$, which can be detected by a
three-Josephson-junction superconducting quantum interference
device (SQUID). The result of measurement can be defined as
$\delta P=\frac{1}{2}|P_{e}-P_{g}|$, where $P_{e(g)}$ denotes the
population of excited (ground) state of phase qubit.  $\delta P=0$
means that the system stands at the delocalized phase, while
$\delta P=0.5$  stands at the localized phase. The $\delta P$ as a
function of the strength $\alpha$, calculated with numerical
renormalization group method, is plotted in Fig.\ref{Fig2}. The
readout technique of phase qubit \cite{capacitance} is achieved by
applying a short bias current pulse $\delta I(t)$ (less than $5$
ns) that adiabatically reduces the well depth $\Delta
U/\hbar\omega_{p}$, so that the first excited state lies very near
the top of the well when the current pulse is at its maximum
value. It has been shown that only the expectation of $\sigma_{z}$
in the phase qubit can be measured; however, the other direction
measurements can be achieved by applying a transformation that
maps the detected eigenvector onto the eigenvector of $\sigma_{z}$
before detecting. The ratio of the tunneling rates for the two
states $|1\rangle$ and $|0\rangle$ is about 200, so the fidelity
of measurement of the phase qubit is about $96\%$ when properly
biased.
%Note that the frame of phase qubit in our setup has been
%rotated, the tomography measurement should be addressed.

Let us briefly introduce the critical behaviors of the model
studied by NRG, which is an efficient way to treat the spin-boson
model with a broad and continuous spectrum of
energies\cite{NRG,Bulla}.
 The NRG method starts with a logarithmic
discretization of the bosonic bath in intervals
$[\Lambda^{-(n+1)}\omega_{c},\Lambda^{-n}\omega_{c}]
(n=0,1,2,\ldots)$, where $\Lambda>1$ is called as the NRG
discretization parameter.  We take the surface plasma frequency as
the cutoff energy ($\omega_{c}=10^{14}$ Hz) and energy unit. After
a sequence of transformations, the discretized model is mapped
onto a semi-infinite chain with the spin representing the first
site of the chain. The spin-boson model in the semi-infinite chain
form \cite{Bulla} is diagonalized iteratively, starting from the
spin site and successively adding degrees of freedom to the chain.
The exponentially growing Hilbert space in the iterative process
is truncated by keeping a certain fraction of the lowest-lying
many particle states. Due to the logarithmic discretization, the
hopping parameters between neighboring sites fall off
exponentially, going along the chain corresponds to accessing
decreasing energy scales in the calculation. From the energy of
the first excited state, the properties of the total system can be
presented according to the parameters, i.e.,
$\Delta,\alpha,\epsilon$.
 It should be mentioned that, when the
controlling parameters are close to the critical values, the
numerical computation would occur small deviation by NRG. The
result of the critical value that obtained directly by numerical
calculation is less than practical one, which is analogous to that
$\epsilon$ is finite small value. In order to illuminate this
errors, we consider the cases of finite small values $\epsilon$ to
show the effect on $\alpha$, as described in Fig.\ref{Fig2}.
%, which calculated by NRG method
%The numerical result also can demonstrate the QPT from delocalized
%phase to localized phase in principle. However, the deviation
%doesn't happen in the practical experiments. So the result of
%experiments will show the phase transition at the proper critical
%points.

\begin{figure}
\includegraphics[width=5.5cm, angle=90]{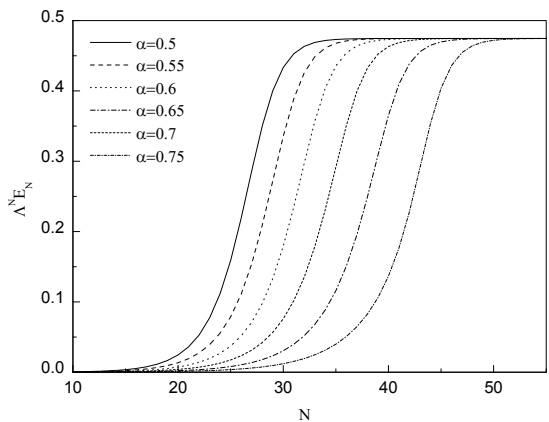}
\caption{Scaling of the flow of the many-particle levels
$E_{N}(r)$ for fixed $s=1,\epsilon=0$, and $\Delta$ tuned in the
regime $2.5\times10^{-5}\sim3.75\times10^{-5}$, and the
corresponding $\alpha$ is changed in $0.5\sim0.75$. The NRG
parameters are $N_{s}=100, N_{b}=6$, and $\Lambda=2.0$.}
\label{Fig3}
\end{figure}

We expect to observe scaling behavior in all physical properties
for $\alpha(\Delta)\rightarrow\alpha_{c}(\Delta_{c})$. An example
is shown in Fig.\ref{Fig3} for various values of $\alpha(\Delta)$.
In this way we can easily determine the critical value
$\alpha_{c}$ from the relation
$T^{*}=const.\times\Lambda^{-N^{\ast}}\propto\Delta^{1/(\alpha_{c}-\alpha)}$
($N^{\ast}$ is the value of $N$ where the first excited state
reaches the value $\Lambda^{N}E_{N}=0.3$, the NRG discretization
parameter is $\Lambda=2.0$ in the paper). From the sets of data
$\{N,\alpha\}$ that approaching critical point in Fig.\ref{Fig3},
the critical value can be determined as $\alpha_{c}=1.093$ via
extrapolation, which is consistent with the exact value
$\alpha_c=1$ in the small $\Delta$ limit \cite{Leggett}. The
critical value separate two different phases, which can be
observed directly by measuring the populations of the qubit
states.

Before ending the paper, we make two additional remarks. If the
coupling capacitance $C_{0}$ can be tuned, we can also detect the
quantum KT phase transition through modifying $C_0$ but fixing
$\Delta$. On the other hand, we just discussed the Ohmic spectra
in the paper since the resistances of the transmission line are
neglected; however, the sub-Ohmic spin-boson model \cite{Tong} can
be designed if a RC-dominate transmission line is considered. In
this case, the critical properties of the sub-Ohmic spin-boson
model can be observed in a slightly expanded model.

%\section{Conclusion}

In summary, we have presented a scheme to realize the spin-boson
model by using a phase qubit coupled to a semi-infinite
transmission line. By tuning the bias current or the coupling
capacitance, the notable KT phase transition from the delocalized
phase to the localized phase can be directly observed through
measuring the  states of the phase qubit. We have also estimated
the required experimental parameters by using numerical
renormalization group.

%\section{Acknowledgements}
This work was supported by the NSFC (Nos. 10974059 and 11004065),
the NSF of Guangdong province, the State Key Program for Basic
Research of China (Nos. 2007CB925204 and 2011CB922104), and the
RGC of Hong Kong (No. HKU7049/07P).

\end{document}